\documentclass{kluwer}

\usepackage{epsfig}

\def\arcsec{$''$}

\def\farcs{\hbox{$.\!\!^{\prime\prime}$}}
\def\kms{$\rm km\;s^{-1}$}
\def\deg{$^\circ$}

\def\etal{{\em et al.\, } }
\def\ha{H$\alpha$}

\begin{document}

\begin{article}

\begin{opening}

%%%%%%%%%% TITOLO E AUTORI %%%%%%%%%%%%%%%%%%%%%%%%%%%%%%

\title{Correlation between kinematics and Hubble Sequence in disk galaxies?}

\author{ J.C. \surname{Vega Beltr\'an}\email{jvega@pd.astro.it} \\ }
\author{E. \surname{Pignatelli}\email{pignatelli@pd.astro.it}}
\institute{Osservatorio Astronomico di Padova, vicolo dell'Osservatorio 5, 
I 35122 Padova, Italy}

\author{ W.W. \surname{Zeilinger}\email{wzeil@magellan.astro.univie.ac.at} }
\institute{Institut f\"ur Astronomie, Universit\"at Wien, Wien, Austria }

\author{A. \surname{Pizzella}\email{apizzell@eso.org} }
\institute{European Southern Observatory, Lasilla 19001, Santiago 19, Chile}

\author{ E.M. \surname{Corsini}\email{corsini@pd.astro.it} \\ }
\author{F. \surname{Bertola}\email{bertola@pd.astro.it} }
\institute{Dipartimento di Astronomia, Universit{\`a} di Padova, 
vicolo dell'Osservatorio 5, I-35122 Padova, Italy }

\author{J. \surname{Beckman}\email{jeb@ll.iac}}
\institute{Instituto de Astrof\'\i sica de Canarias, La Laguna, Spain}

%%%%%%%%%%%%%%%%% ABSTRACT %%%%%%%%%%%%%%%%%%%%%%%%%%%%%%%%%%%%

\begin{abstract}

We present a comparison between ionized gas and stellar kinematics for
a sample of 5 early-to-intermediate disc galaxies.  We measured the
major axis $V$ and $\sigma$ radial profiles for both gas and stars,
and the $h_3$ and $h_4$ radial profiles of the stars. We also derived
from the $R$-band surface photometry of each galaxy the light
contribution of their bulges and discs.

In order to investigate the differences between the velocity fields of
the sample galaxies we adopted the self-consistent dynamical model by
Pignatelli \& Galletta (1999), which takes into account the asymmetric
drift effects, the projection effects along the line-of-sight and the
non-Gaussian shape of the line profiles due to the presence of
different components with distinct dynamical behavior. We find for the
stellar component a sizeable asymmetric drift effect in the inner
regions of all the sample galaxies, as it results by comparing their
stellar rotation curves with the circular velocity predicted by the
models.

The galaxy sample is not wide enough to draw general conclusions.
However, we have found a possible correlation between the presence of
slowly-rising gas rotation curves and the ratio of the bulge/disc half
luminosity radii, while there is no obvious correlation with the key
parameter represented by the morphological classification, namely the
bulge/disc luminosity ratio.  Systems with a diffuse dynamically hot
component (bulge or lens) with a scale length comparable to that of
the disc are characterized by slowly-rising gas rotation curves. On
the other hand, in systems with a small bulge the gas follows almost
circular motions, regardless of the luminosity of the bulge itself.
We noticed a similar behavior also in the gas and stellar kinematics
of the two early-type spiral galaxies modeled by Corsini et
al. (1998).

\end{abstract}

\end{opening}

%%%%%%%%%%%%%%%% INIZIA l'ARTICOLO %%%%%%%%%%%%%%%%%%%%%%%%%%%%%%%%%%%%%%%

\section{Introduction}

Evidence has been presented (Bertola \etal 1995) that in several S0
galaxies the gas in the inner regions shows broad velocity dispersion
profiles. In addition, in  the centers of certain early-type galaxies
it was discovered (Fillmore \etal 1986; Kent \etal 1988) that the gas
rotation curves fall down the circular velocities predicted by constant
M/L models.  Both gas phenomena: the ``slowly-rising'' rotation curve
and the high velocity dispersion, can be well interpreted  as signs
of gas recently expelled by the stars of the bulge but not yet heated
to the virial temperature of the galaxy (i.e. {\it
pressure-supported}).  In the case of the late-type spirals the picture
is clearer: both the gas and the stars show essentially  circular
motions and  constant velocity dispersion profiles.

It has been shown for Sa galaxies (Corsini \etal 1998) that the
application of dynamical models can tell us a great deal about the
structure and the kinematics of galaxies: understanding of the
dynamics for the different stellar components, deviation of the gas
from circular motions and the presence of dark matter. Here
we present a comparison between the gas and stellar kinematics for a
sample of 5 disc galaxies (see Table~\ref{tab:models_sample}) of early
and intermediate Hubble types (Sa, Sb).

\begin{table}[H]
\begin{center}
\begin{footnotesize}
\begin{tabular*}{\maxfloatwidth}{llrcccrr}
\hline

\multicolumn{1}{c}{object} &
\multicolumn{1}{c}{type} &
\multicolumn{1}{c}{$B_T$} &
\multicolumn{1}{c}{$i$} &
\multicolumn{1}{c}{$V_0$} &
\multicolumn{1}{c}{$D$} &
\multicolumn{1}{c}{scale} &
\multicolumn{1}{c}{$R_{25}$} 
						 \\
\multicolumn{1}{c}{\normalsize [name]} &
\multicolumn{1}{c}{\normalsize [RC3]} &
\multicolumn{1}{c}{\normalsize [mag]} &
\multicolumn{1}{c}{\normalsize [\deg]} &
\multicolumn{1}{c}{\normalsize [\kms]} &
\multicolumn{1}{c}{\normalsize [Mpc]} &
\multicolumn{1}{c}{\normalsize [pc/$''$]} &
\multicolumn{1}{c}{\normalsize [$''$]} 
						\\
\multicolumn{1}{c}{({1})} &
\multicolumn{1}{c}{({2})} &
\multicolumn{1}{c}{({3})} &
\multicolumn{1}{c}{({4})} &
\multicolumn{1}{c}{({5})} &
\multicolumn{1}{c}{({6})} &
\multicolumn{1}{c}{({7})} &
\multicolumn{1}{c}{({8})} 
						\\
\hline
NGC  772 & .SAS3.. & 11.09 &  53  &$2618$& 34.7 &  168  & 217  \\
NGC  980 & .L..... & \dots &  62  &$5936$& 79.1 &  384  &  51  \\
NGC 3898 & .SAS2.. & 11.60 &  46  &$1283$& 17.1 &   83  & 131  \\
NGC 5064 & PSA.2*. & 12.69 &  65  &$2750$& 36.7 &  178  &  74  \\
NGC 7782 & .SAS3.. & 13.08 &  62  &$5611$& 74.8 &  363  &  72  \\
\hline
\end{tabular*} 
\end{footnotesize}
\caption{Parameters of the Sample Galaxies. 
Col.({3}): total observed blue magnitude from RC3 except for NGC~5064 (RSA).
Col.({4}): inclination from NGB (Tully 1988) except for NGC~980 and NGC~7782 
	   from RC3 (inferred from R$_{25}$).
Col.({5}): systemic velocity derived from the center of symmetry
	   of the gas rotation curve  and corrected for local motion. 
Col.({6}): distance obtained from $V_0/H_0$ with $H_0=75$ \kms\ Mpc$^{-1}$.  
Col.({8}): radius of the 25 $B-$mag arcsec$^{-2}$ isophote from RC3. 
}
\label{tab:models_sample}
\end{center}
\end{table}

\vspace{-1cm}
The study of the Sb galaxies allows us to see whether non circular
motions are present in the inner regions for both gas and stars, or if
we are facing systems with intermediate kinematic properties between
early and late-type objects as might be inferred from the photometric
properties. 

\section{Observations and data deduction}

Spectroscopic observations were carried out at ESO, La Silla (Chile) (ESO
N.~60.A-0800), at the MMT operated on Tucson
(Arizona) and at the INT operated on the island  La Palma
by the Royal Greenwich Observatory in the Spanish Observatorio del
Roque de los Muchachos of the Instituto de Astrof\'{\i}ca de Canarias
(IAC), Tenerife (Spain). 

Photometric observation were carried out at the IAC80 operated on the
island Tenerife by the IAC in the Spanish Observatorio of
Iza\~na, at the VATT operated on Tucson, and at  ESO, La Silla (Chile) (ESO
N.~60.A-0800).

The standard spectral reduction was performed by using the ESO-MIDAS
package.  The stellar kinematics was measured from the absorption
lines present on each spectrum using the Fourier Correlation Quotient
Method (Bender 1990) as applied by Bender, Saglia \& Gerhard (1994).
The ionized gas velocities and velocity dispersions were derived by
means of MIDAS package ALICE. 
The photometric data reduction was carried out using standard IRAF
routines.

\section{The model}

In order to investigate the different behavior of the gas and the
stellar kinematics a self-consistent dynamical model (Pignatelli \&
Galletta, 1999; see also Pignatelli \& Galletta, elsewhere in this
volume) is used to reproduce the data.

The model has $4n+1$ free parameters, where $n$ is the number of
adopted components: namely the luminosity, scale length,
mass-luminosity ratio and flattening $\{L_{Tot},\ r_e,\ M/L,\ b/a \}$
of each component plus the inclination angle of the galaxy. In
principle, all of these parameters can be constrained - within some
confidence limits - by the photometry, with the exception of the M/L
ratios, which have to be derived from the kinematics.

%We can then
%integrate the Jeans equations, obtaining a self-consistent model of
%the rotational velocity and velocity dispersion profiles for the
%different components, including the asymmetric drift effects.

In order to compare the observed data with the predictions of the
model, we also need to reproduce the deviations of the LOSVD from a
pure Gaussian shape.  We adopted the usual parameterization in terms of
a Gauss-Hermite expansion series (van der Marel and Franx 1993;
Gerhard 1993).

%In fact, in the regions where the bulge and disc
%luminosities are comparable we expect that the superposition of the
%rapid rotation of the disc on the slower rotation of the other
%components will produce a non-Gaussian, and sometimes even 2-peaked,
%LOSVD even assuming that each component has a Gaussian velocity
%distribution.  

Once the fit of the stellar kinematics has been performed, and the
overall potential of the galaxy is known, one can directly derive the
circular velocity $V_c$.  By superimposing the $V_c$ inferred in this
way (and corrected for inclination) on the observed gas rotational
velocity, we can immediately notice any deviation from purely circular
motions.

\section{Results}

\vspace{-0.5cm}

For shortness reason, we give in detail the results for only 2 of the
5 galaxies studied, and briefly summarize the other ones.  We refer to
the complete paper (Vega \etal 1999) for details. 

\begin{figure}[H]
\centerline{{\psfig{figure=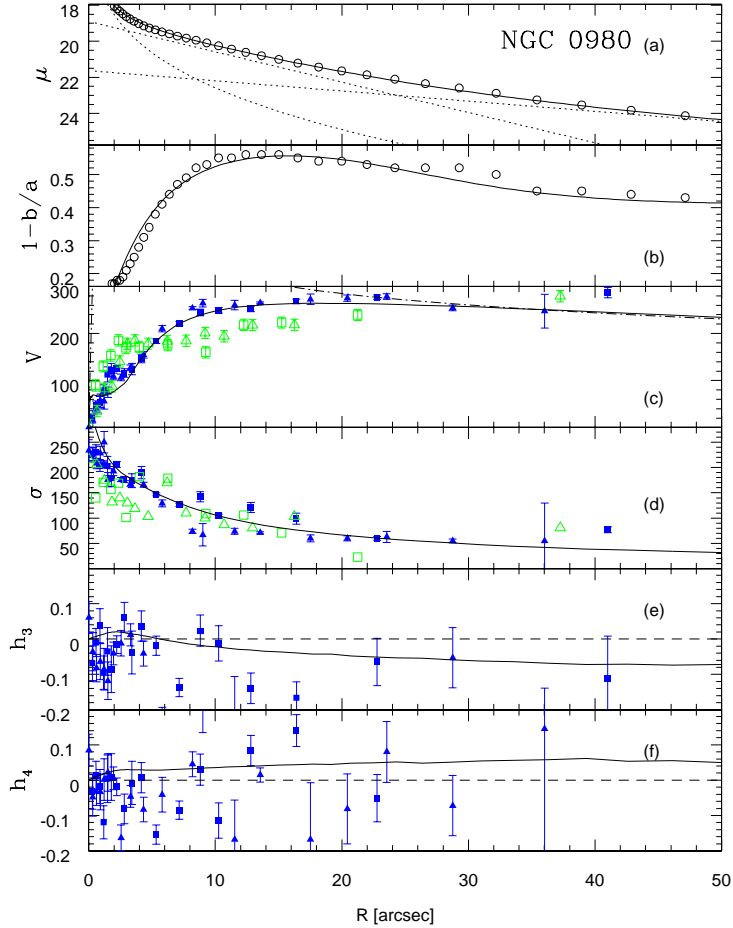,width=10cm}}}
\caption[Photometric and dynamical features for  NGC~980]
{Photometric and dynamical features for NGC~980 with their
respective best fit curves obtained from our models. 
(a): r-band surface brightness (open circles) as a function of
    radius along the major axis.  We also plot the photometric
    components that we found (dashed lines) and the sum of these
    components (solid line). 
({b}): ellipticity (open circles)
    associated to the upper r-band surface brightness and the best fit
    curve (solid line) founded.
({c}): observed stellar velocity (filled symbols) with its
    associated model (solid line), and the \ha\ observed gas
    kinematics (open symbols). The {\it dot-dashed} line is the circular
    velocity inferred from the model. 
({d}): the same as (c) but for the velocity dispersion.  
({e}): same as (c) for the h$_3$ coefficients of the Gauss-Hermite
    expansion of the line profile of the stars. 
({f}): same as (c) for the h$_4$ coefficients of the Gauss-Hermite
    expansion of the line profile of the stars.
In windows ({c}) to
({f}) the squares symbols and the triangle symbols represent data
derived for the approaching NE side and for the receding SW side
respectively. In window ({c}) due to the high values of the predicted
circular velocity its fitting curve is only visible for the outer
regions. The plot of the entire curve would only produce a loss of the
velocity curve details. }
\label{fig:models_0980}
\end{figure}

{\bf NGC~980:}
The photometric model adopted for this lenticular galaxy is that of an
almost edge-on disc component embedded in a lens much more extended
than the disc itself, together with a very compact, luminous bulge in
the inner 2\arcsec\ (see Fig.~\ref{fig:models_0980}).  The
superposition of these 3 components accounts for the peak in
ellipticity at 10\arcsec, and the subsequent slow decrease from
$\varepsilon=0.6$ to $\varepsilon=0.3$.

The most interesting feature of this galaxy is surely the gas-stars
kinematic difference.  From 3\arcsec\ to 10\arcsec\ gas and stars
appear to have quite a different behavior, with the gas raising
linearly in velocity, while the stars do present a plateau.  As a
consequence, the gas is rotating faster at 3\arcsec\ than the stars,
and {\em slower} than the stars at 10\arcsec. On the other hand, both
the components are following the same velocity dispersion curve,
marking the likely association of the gas with a hot component.  This
difference between the gas and stellar rotational velocity is unlikely
to be due only to absorption effect, and could be due to a
misalignment between the gas and the stellar component, suggested also
by the slow change of about 10\deg\ in the position angle of the {\it
R}-band isophotes.

At 2\arcsec\ it is also visible a small peak in the velocity, which is
not evident in the gas rotation curve, and that we interpreted as the
effect of the rotation of a massive central flattened distribution of
matter, also responsible for the sharp increase in the velocity
dispersion in the inner region.  Finally, the presence of a lens is
inferred from the slow decrease of the velocity dispersion, much
slower than would be expected from a small bulge; as well as from the
stellar LOSVD, strongly non-Gaussian at r=8\arcsec.

\begin{figure}[H]
\centerline{{\psfig{figure=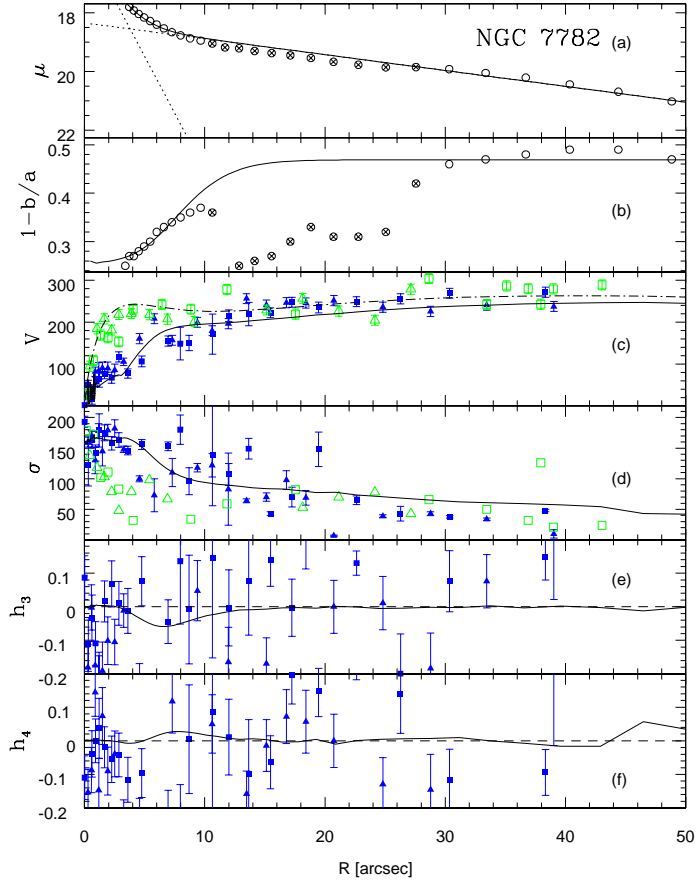,width=9.5cm}}}
\caption[Photometric and dynamical features for  NGC~7782]
{Same as Fig.~\ref{fig:models_0980} for NGC~7782. In windows ({a}) and
({b}) the circle symbols filled with crosses are not considered to find 
the best
photometric decomposition due to the presence of strong spiral arms.
For windows ({c}) to ({f}) the squares symbols and the
triangle symbols represent data derived for the approaching NW side
and for the receding SE side respectively. }
\label{fig:models_7782}
\end{figure}

{\bf NGC~7782:} This Sb galaxy 
looks like the prototype of a bulge+disc system, with the
two components almost decoupled from each other. From both the
kinematic and the photometric data we can distinguish two different
regions with very different behaviors.  In the inner 5\arcsec\
region, dominated by an almost spherical bulge, the stellar velocity
dispersion shows a plateau at $\sim$200 \kms\ and the rotational
velocity is less than half the value expected from the circular
velocity, in agreement with the asymmetric drift effect calculated by
the model (see Fig.~\ref{fig:models_7782}).
On the other hand the gas shows strictly
circular motions with low velocity dispersion. In the outer region,
well approximated by an exponential disc, both 
the gas and the stars appear to rotate close to the circular velocity.

Both an exponential and a r$^{1/4}$ law has been adopted for the
potential models of the bulge. While the photometry itself can not distinguish

\noindent between the two hypothesis, we found that the exponential bulge gives
a better agreement with the kinematic data.

{\bf NGC 772:} The gas seems to be pressure supported in the inner
6\arcsec ; after that the rotational velocity of the gas reaches the
expected circular velocity and the velocity dispersion falls to a low
constant value ($\sim$ 50 \kms).

{\bf NGC~3898:} The stellar kinematics is well fitted by a 3-component
model, including an inner flat disc coexisting with a spherical bulge
of similar scale length. The gas rotational velocity is well
approximated by the model circular velocity for r$>$8\arcsec, showing
no asymmetric drift effect, and is observed to follow a flat velocity
curve out to 100\arcsec. In the inner 8\arcsec\ there is a strong
absorption of the \ha\ line, that prevent us from obtaining realist of
the velocity and the velocity dispersion.

{\bf NGC~5064:} Apart from a slight oscillation in the ellipticity and the
position angle, likely due to the spiral arms, this galaxy does not
show evidence for triaxial structures or misalignments in the
different stellar components. The position angle is fairly constant
and there is no sign of boxy structures. Nevertheless, with the
hypothesis of our models, we found that no set of parameters was able
to reproduce the data. A possible explanation could be that the
observed value of the rotational velocity is the mean of the
superposition of different components in the central 20\arcsec\
region, with different characteristic values of the rotational
velocity and/or directions of the angular momentum. This possibility
is supported also by the $h_3$, $h_4$ profiles which mark a strongly
non-Gaussian shape of the line-of-sight velocity profiles in this
region. Even inner stellar counter-rotation can not be ruled out at
this moment.

\vspace{-0.3cm}

\section{Discussion}

We analyzed the photometry and the kinematics of 5 disc galaxies (one
is classified as lenticular, two as Sa and two as Sb by the RSA
catalog). For 4 out of 5 galaxies, we found that an axisymmetric
dynamical model can fully account for the observed stellar kinematic
features, including the general behavior of the h$_3$, h$_4$
profiles. For the case of NGC~5064, models with decoupled angular
momenta may be needed.

In {Figs.~\ref{fig:models_0980}-\ref{fig:models_7782}} we present the
photometric and spectroscopic data for two of these galaxies, together
with the best fit models obtained. The parameters obtained from the
models are presented in Table~\ref{tab:models_param1}.
A sizable asymmetric drift effect is present in the inner regions of
all these galaxies, as we can see by comparing the stellar rotational
curves with the circular velocity predicted by the models.

The gas components show a very wide range of kinematic properties.
Two of the sample galaxies (NGC~5064 and NGC~7782) show almost
circular motions even very near the center; two present slowly rising
rotation curves (NGC~772 and NGC~3898) and one (NGC~980) even appears
to have a gas rotational curve {\em lower} in velocity than that of
the stellar component.

The number of galaxies belonging to the sample is not large enough to
draw general conclusions. However we have found a possible correlation
between the presence of slowly-rising gas rotation curves and the
ratio of the bulge/disc half luminosity radii, while there is no
obvious correlation with the key parameter represented by the
morphological classification, namely the bulge/disc luminosity ratio.
Systems with a diffuse dynamically hot component (bulge or lens), with
scale length comparable to that of the disc are characterized by
slowly rising gas rotation curves. On the other hand, in systems with
a small bulge the gas follows almost circular motions, regardless of
the luminosity of the bulge itself.  We notice similar behavior also
in the gas and stellar kinematics of two Sa galaxies (NGC~2179 and
NGC~2775) modeled by Corsini \etal  (1998). The application of the
methods of analysis used in this contribution to a wider sample of
galaxies could help us to confirm this picture.

\begin{table}[H]
\caption{Parameters from the Dynamical Models (I). The index $b$ and $d$ refers, respectively, to the bulge and disk component. The $d'$ index to the (eventual) third component.} 
\begin{center}
\begin{footnotesize}
\begin{tabular*}{\maxfloatwidth}{lrrrcccrrr}
\hline
&
\multicolumn{3}{c}{scale radius} 	&
\multicolumn{3}{c}{axial ratio}  	&
\multicolumn{3}{c}{Mass [$10^{10} M_\odot$]} 		
	  \\ \lcline{2-4} \cline{5-7}  \rcline{8-10}

\multicolumn{1}{c}{object} 	 	&
\multicolumn{1}{c}{r$_b$} 		&
\multicolumn{1}{c}{r$_d$} 		&
\multicolumn{1}{c}{r$_{d^\prime}$} 	&
\multicolumn{1}{c}{(b/a)$_b$} 		&
\multicolumn{1}{c}{(b/a)$_d$} 		&
\multicolumn{1}{c}{(b/a)$_{d^\prime}$} 	&
\multicolumn{1}{c}{M$_b$} 		&
\multicolumn{1}{c}{M$_d$} 		&
\multicolumn{1}{c}{M$_{d^\prime}$} 	
					\\
\hline

NGC  772 &19\farcs 0&20\farcs0 & ---     & 0.80 & 0.1 & --- & $6.5 $ & $9.9$ &  ---   \\
NGC  980 &2\farcs 1 &6\farcs 5 &19\farcs2& 0.98 & 0.2 & 0.6 & $24.0$ & $4.9$ & $10.0$   \\
NGC 3898 &9\farcs 5 &6\farcs 6 &57\farcs0& 1.00 & 0.2 & 0.3 & $3.7$  & $0.5$ & $3.2$   \\
NGC 7782 &1\farcs 4 &20\farcs 1&  ---    & 0.75 & 0.1 & --- &$6.0$  & $36.0$ &  ---   \\
\hline
\end{tabular*} 
\end{footnotesize}
\end{center}
\label{tab:models_param1}
\end{table}

\end{article}

\end{document}